\documentclass[preprint,showpacs,preprintnumbers,amsmath,amssymb,floatfix]{revtex4}
\usepackage{graphicx}
\usepackage{supertabular}
\usepackage{epsfig}
\usepackage{dcolumn}
\usepackage{bm}
\usepackage{wrapfig} 
\usepackage{amsmath}
\usepackage{amsfonts}
\usepackage{mathrsfs}

\begin{document} 

\title{Microscopic-Macroscopic Approach for Binding Energies with the
Wigner-Kirkwood Method - II}
\author{A. Bhagwat$^{1}$, X. Vi\~nas$^{2}$, M. Centelles$^{2}$, 
P. Schuck$^{3,4}$ and R. Wyss$^{5}$}
\affiliation{$^{1}$UM-DAE Centre for Excellence in Basic Sciences, 
Mumbai 400 098, India \\
$^{2}$Departament d'Estructura i Constituents de la Mat\`eria
and Institut de Ci\`encies del Cosmos, Facultat de F\'{\i}sica,
Universitat de Barcelona, Diagonal {\sl 645}, {\sl E-08028} Barcelona,
Spain \\
$^{3}$Institut de Physique Nucl\'eaire, IN2P3-CNRS, Universit\'e Paris-Sud, 
F-91406 Orsay-C\'edex, France \\ 
$^{4}$ Laboratoire de Physique et Mod\'elisation des Milieux Condens\'es,
CNRS and Universit\'e Joseph Fourier, 25 Avenue des Martyrs, Bo\^{i}te Postale 
166, F-38042 Grenoble Cedex 9, France  \\ 
$^{5}$KTH (Royal Institute of Technology), Alba Nova University Center, 
Department of Nuclear Physics, S-10691 Stockholm, Sweden}
\date{\today}

\begin{abstract}
The binding energies of deformed even-even nuclei have been analysed within the 
framework of a recently proposed microscopic-macroscopic model. We have used the semiclassical 
Wigner - Kirkwood $\hbar$ expansion up to fourth - order, instead of the usual 
Strutinsky averaging scheme, to compute the shells corrections in a deformed 
Woods - Saxon potential including the spin-orbit contribution. For a large set 
of 561 even-even nuclei with $Z\ge 8$ and $N\ge 8$, we find an {\it rms} 
deviation from the experiment of 610 keV in binding energies, comparable to the 
one found for the same set of nuclei using the FRDM of M\"oller and Nix (656 keV). As 
applications of our model, we explore its predictive power near the proton and 
neutron drip lines as well as in the superheavy mass region. Next, we 
systematically explore the fourth - order Wigner - Kirkwood corrections to the 
smooth part of the energy. It is found that the ratio of the fourth - order to 
the second - order corrections behaves in a very regular manner as a function 
of the asymmetry parameter $I=(N-Z)/A$. This allows to absorb the fourth - order
corrections into the second - order contributions to the binding energy, which 
enables to simplify and speed up the calculation of deformed nuclei. 
\end{abstract}
\maketitle

\section{Introduction}
The models of nuclear masses are continuously challenged by the advances in 
experimental techniques which nowadays are extending the nuclear chart to 
previously unexplored regions of exotic isotopes and superheavy elements. The 
theoretical description of nuclear masses takes place primarily along two main 
approaches. On the one hand, in the microscopic nuclear models, the nuclear 
binding energy is obtained from calculations with energy density functionals 
based on effective nuclear interactions \cite{BEN.03,LUN.03,STO.07}. In the 
microscopic-macroscopic (mic-mac) models \cite{MS.77,POM.03,LUN.03}, the 
nuclear binding energy is obtained as the sum of a part that varies smoothly 
with the number of nucleons plus an oscillatory correction originated by the 
quantum effects. The smooth part of the mic-mac models is obtained from a 
liquid-drop model approach, whereas the shell correction is usually evaluated 
by the Strutinsky averaging method in an external potential well.

In our previous works \cite{BHA.10,BHA.10a}, we have demonstrated that the 
Strutinsky average can be replaced by the semiclassical energy computed by 
means of the Wigner - Kirkwood (WK) $\hbar$ expansion of the one - body partition function
\cite{WIG.32,KIR.33,JEN.75,RS.80,BRA.85,BRA.97,CEN.06,CEN.07}, in order to 
evaluate the shell corrections of a system of $N$ neutrons and $Z$ protons at 
zero temperature in an external potential. There are some reasons supporting 
this choice as we have discussed in Ref. \cite{BHA.10}. On the one hand, it has 
been shown that the Strutinsky level density is an approximation to the WK level
density \cite{AZI.06}. On the other hand, the WK $\hbar$-expansion of the 
density matrix has a variational content and it is possible to establish a 
variational theory based on a strict $\hbar$-expansion \cite{CEN.07,XV.93}. We 
shall point out that the WK expansion is also well suited to deal with nuclei 
close the drip lines. Although the WK level density exhibits a well known 
$\varepsilon^{-1/2}$ divergence as $\varepsilon \to 0$ for a potential that 
vanishes at large distances, integrated moments of the level density, such 
as the energy and the accumulated level density, are well behaved in the 
$\varepsilon \to 0$ limit as it has been demonstrated in Ref.\cite{CEN.07}. It 
has been shown that these shell corrections, along with a simple six parameter 
liquid drop formula, yield a good description of ground - state masses of 
spherical nuclei spanning the entire periodic table \cite{BHA.10}. The model 
has also been applied to calculate the binding energies of few deformed nuclei, 
with a good degree of success \cite{BHA.10a}. In the present work, we extend 
the work reported earlier \cite{BHA.10} to the deformed nuclei and explore 
the predictions of the model in exotic scenarios such as drip line nuclei
and the superheavy region. In this work, we mainly restrict our attention to the even - even 
nuclei.

One of the important conclusions of Ref. \cite{BHA.10} is that in this model it is necessary 
to carry out the WK expansion up to the fourth - order in $\hbar$ to obtain 
accurate shell corrections, which implies that in this case one needs to work out derivatives 
of the single particle potentials (nuclear potential, Coulomb potential as well 
as the spin - orbit potential) up to the fourth - order, which is a rather 
cumbersome task. Therefore, this gives rise to an interesting and important 
question: can the effects of the fourth - order corrections to the binding 
energy be absorbed into the second - order ones? This question is important 
from theoretical as well as practical point of view. Theoretically, this would 
imply that the WK series has been partially re-summed, whereas from a practical 
point of view, it implies that it is sufficient to expand the one - body partition
function up to second - order in $\hbar$ to obtain shell corrections with 
comparable accuracy.

The absorption, if possible, would imply that there is a factor (we denote the 
factor by $\alpha$), which may be a function of mass number, charge number, 
neutron number or combinations thereof, defined as
\begin{eqnarray}
\alpha ~=~ 1~+~\frac{E\left(\hbar^4\right)}{E\left(\hbar^2\right)}
\label{alpha}
\end{eqnarray}
such that 
\begin{eqnarray}
E\left(\hbar^2\right)~+~E\left(\hbar^4\right)~=~\alpha E\left(\hbar^2\right)
\end{eqnarray}
where, $E\left(\hbar^2\right)$ and $E\left(\hbar^4\right)$, respectively, are
second and fourth - order WK corrections to energy. This is an important issue 
discussed in the present article. 

We summarise the essential details of the semiclassical Wigner - Kirkwood 
expansion of the one - body partition function in the second section. The detailed 
results and their analysis forms the subject matter of the third section. The 
parameters of the macroscopic part of our mic-mac model, which also includes 
curvature correction \cite{POM.03} and the Wigner term \cite{POM.03}, have been 
obtained by minimizing the $\chi^2$ value of the energies using a selected set 
of 561 even-even deformed and spherical nuclei. The ability of this mic-mac model
to describe nuclei in the exotic scenarios is explored in section 4. On the one 
hand, masses of very proton rich nuclei, measured recently \cite{XLT.11}, are 
compared with the predictions of our model. On the other hand, the upper limit 
of the outer crust in neutron stars is studied, which involves nuclei near the 
neutron drip line. Finally, we explore the superheavy region, and compare the 
theoretical alpha decay $Q$ values and the corresponding half lives with the 
experimental values \cite{OGA.07}. The systematic investigation of the 
absorption factor $\alpha$ as defined above is contained in the fifth section. 
The summary and conclusions are given in the last section.

\section{Formulation}
For a system of $N$ non interacting Fermions at zero temperature in a given 
external potential, the quantal one - body partition function is given by:
\begin{eqnarray}
Z\left(\beta\right)~=~\mathrm{Tr}\left(\exp{(-\beta\hat{H})}\right)~.
\label{Z_qm}
\end{eqnarray} 
The Hamiltonian of the system ($\hat{H}$) is expressed as:
\begin{eqnarray} 
\hat{H}~=~\frac{-\hbar^2}{2m} \nabla^2 ~+~V(\vec{r})~+~\hat{V}_{LS}(\vec{r}) \,,
\end{eqnarray} 
with $V(\vec{r})$ being the one-body central potential and
$\hat{V}_{LS}(\vec{r})$ the spin-orbit interaction. The replacement of the 
Hamiltonian in the above equations by the corresponding classical Hamiltonian 
leads to the well - known Thomas - Fermi equations for particle number and 
total energy. The Wigner - Kirkwood semiclassical expansion amounts to 
expansion of the quantal one - body partition function in the powers of 
Planck's constant, $\hbar$, yielding systematic corrections to the Thomas - 
Fermi energy and particle number 
\cite{WIG.32,KIR.33,JEN.75,RS.80,BRA.85,BRA.97}.

As stated before, in this work, we carry out the WK expansion up to the fourth -
order in $\hbar$. With the spin - orbit interaction, the WK expansion of 
the partition function can be written schematically as:
\begin{eqnarray} 
Z_{WK}^{(4)}(\beta)~=~Z^{(4)}(\beta)~+~Z^{(4)}_{SO}(\beta)~.
\end{eqnarray} 
where, $Z^{(4)}(\beta)$ ($Z^{(4)}_{SO}(\beta)$) is the WK partition function 
for the central potential (spin - orbit part). The explicit expressions for 
these partition functions can be found in \cite{JEN.75,BHA.10}.

The level density $g_{WK}$, the particle number $N$ and the energy $E_{WK}$ 
are obtained by appropriate Laplace inversion of the WK partition function, 
as follows:
\begin{eqnarray} 
g_{WK}(\epsilon)~=~{\cal{L}}^{-1}_{\epsilon} Z_{WK}^{(4)}(\beta)~,
\end{eqnarray} 
\begin{eqnarray} 
N~=~{\cal{L}}^{-1}_{\lambda} \left( \frac {Z_{WK}^{(4)}(\beta)} {\beta} \right)
\label{N_qm}
\end{eqnarray} 
and
\begin{eqnarray} 
E_{WK}~=~\lambda N~-~{\cal{L}}^{-1}_{\lambda} \left( \frac {Z_{WK}^{(4)}(\beta)} {\beta ^2} \right) ,
\label{E_qm}
\end{eqnarray} 
Here, $\lambda$ is the chemical potential, determined to ensure the correct 
particle number.

The focus of the present article being the WK energy, we present the explicit 
expressions for the WK energies alone. Following Jennings {\it et al.} 
\cite{JEN.75}, the energy (Eq. (\ref{E_qm})) can be written as:
\begin{eqnarray} 
E_{WK} &=& \lambda N - \left( E_{\hbar^{0}}^{CN} + E_{\hbar^{2}}^{CN} + E_{\hbar^{4}}^{CN} \right)
           - \left( E_{\hbar^{2}}^{SO} + E_{\hbar^{4}}^{SO} \right)
\end{eqnarray} 
where, $E_{\hbar^{k}}^{CN}$ denote the contribution to the energy of the order 
$\hbar^{k}$ arising from Laplace inversion 
${\cal{L}}^{-1}_{\lambda}  \left( Z^{(4)}(\beta)/\beta ^2\right) $. On the 
other hand, $E_{\hbar^{k}}^{SO}$ are corrections to the energy of the order 
$\hbar^{k}$ due to Laplace inversion 
${\cal{L}}^{-1}_{\lambda}  \left( Z^{(4)}_{SO}(\beta)/\beta ^2\right) $.
The explicit expression are as follows (see \cite{BHA.10} for further details):
\begin{eqnarray} 
E_{\hbar^{0}}^{CN} &=& \frac{1}{3\pi^2}\left(\frac{2m}{\hbar^2}\right)^{3/2} \int d\vec{r} \left\{
                 \frac{2}{5}\left(\lambda-V\right)^{5/2} \right\} \Theta \left(\lambda-V\right) \\
E_{\hbar^{2}}^{CN} &=& -\frac{1}{24\pi^2}\left(\frac{2m}{\hbar^2}\right)^{1/2} \int d\vec{r} \left\{
                 \left(\lambda-V\right)^{1/2} \nabla^2 V \right\}\Theta \left(\lambda-V\right) \\
E_{\hbar^{4}}^{CN} &=& - \frac{1}{5760\pi^2} \left(\frac{\hbar^2}{2m}\right)^{1/2} \left[ 
\int d\vec{r} \left(\lambda-V\right)^{-1/2}\left\{ 7\nabla^4 V\right\}  \right. \nonumber \\
  && \left. \hspace{3.00cm} 
    + \frac{1}{2} \int d\vec{r} \left(\lambda-V\right)^{-3/2} \left\{ 5\left(\nabla^2V\right)^2+\nabla^2\left(\nabla V\right)^2 \right\}
          \right] \Theta \left(\lambda-V\right) \\
E_{\hbar^{2}}^{SO} &=&  \frac{\kappa^2}{6\pi^2}\left(\frac{2m}{\hbar^2}\right)^{1/2} \int d\vec{r} \left\{ \left(\lambda-V\right)^{3/2} \left(\nabla f\right)^2 \right\}
    \Theta \left(\lambda-V\right)  \\
E_{\hbar^{4}}^{SO} &=&
      \frac{1}{48\pi^2}\left(\frac{\hbar^2}{2m}\right)^{1/2} \int d\vec{r} \left(\lambda-V\right)^{1/2} 
\left[ \kappa^2 \left\{ \frac{1}{2}\nabla^2 \left( \nabla f\right)^2 - \left(\nabla^2 f\right)^2 
         + \nabla f \cdot \nabla \left(\nabla^2 f\right) \right. \right. \nonumber \\ 
  & & \left.\left. - \frac { \left(\nabla f\right)^2 \nabla^2 V } { 2 \left(\lambda-V\right) } \right\} 
      - 2\kappa^3 \left\{ \left( \nabla f\right)^2 \nabla^2 f - \frac{1}{2} \nabla f \cdot \nabla \left(\nabla f\right)^2 \right\}
      +2\kappa^4 \left( \nabla f\right)^4
\right ] \Theta \left(\lambda-V\right) \nonumber \\
\label{E_WK}
\end{eqnarray} 
In these expressions, $V$ is the mean field, $f$ is the spin - orbit form factor,
$\kappa$ is the strength of spin - orbit interaction, and $\lambda$ 
is the chemical potential.

The shell corrections, which are the difference between the quantum 
mechanical and the corresponding averaged energies, can now be obtained by 
subtracting $E_{WK}$ from the quantum mechanical energy.
For our calculations we choose a Woods-Saxon potential as mean field 
and a suitable Woods - Saxon form factor in the spin - orbit sector. These
potentials are generalised for taking into account deformation effects 
and their corresponding parameters are given in Ref. \cite{BHA.10}. 
The Coulomb potential has been obtained by folding the proton density 
distribution with the Coulomb interaction \cite{BHA.10}.
In the microscopic part we have also included pairing correlations using
the Lipkin - Nogami scheme \cite{LIP.60,NOG.64,HCP.73}, as described in 
details in Ref. \cite{BHA.10}.

\section{Calculation of binding energies}
In the present work, we generalise the liquid drop formula
employed in \cite{BHA.10} by adding a deformation dependent curvature energy
term and the Wigner term. The curvature energy term is found to be 
important in improving the agreement achieved between calculations and 
the corresponding experimental binding energies \cite{POM.03}. The Wigner term
is expected to be important for light nuclei as well as to describe nuclei 
close to the proton drip line. Therefore, the modified liquid drop formula 
used in this work reads:
\begin{eqnarray} 
E_{LDM}&=&a_v\left[1~+~\frac{4k_v}{A^2}~T_z\left(T_z~+~1\right)\right]A 
      ~+~ a_s\left[1~+~\frac{4k_s}{A^2}~T_z\left(T_z~+~1\right)\right]A^{2/3} \nonumber \\
&+& a_{cur}\left[1~+~\frac{4k_{cur}}{A^2}~T_z\left(T_z~+~1\right)\right] A^{1/3} 
       ~+~ \frac{3Z^2e^2}{5r_0A^{1/3}}~+~\frac{C_4Z^2}{A} ~+~ E_{W}\,,
\label{LDM}
\end{eqnarray} 
where the terms respectively represent: volume energy, surface energy,
curvature energy, Coulomb energy, correction to Coulomb energy due to 
surface diffuseness of charge distribution and the Wigner energy. The 
coefficients $a_v$, $a_s$, $a_{cur}$, $k_v$, $k_s$, 
$k_{cur}$, $r_0$ and $C_4$ are free parameters; $T_z$ is the third 
component of isospin, and $e$ is the electronic charge.

Several parametrisations of the Wigner term are available in the literature
(see, for example, \cite{POM.03,LUN.03,MN.95}). Here, we adopt the following 
ansatz for the Wigner term with a cut off on charge and mass numbers:
\begin{eqnarray} 
E_{W}~=~w_{1}\exp\left\{-w_{2}\left| \frac{N-Z}{A}\right |\right\} \Theta\left(Z-20\right) 
\Theta\left(A-40\right)
\end{eqnarray} 
where, $w_1$ and $w_2$ are free parameters. The cut offs on charge and mass 
numbers have been introduced since it is expected that the Wigner term will 
make significant contributions for nuclei with low masses.

The Coulomb, surface and curvature terms appearing in the liquid drop formula, 
as defined above in Eq.(\ref{LDM}), need to be modified for the deformed shapes.
In particular, the Coulomb term is multiplied by
\begin{eqnarray}
{\cal{F}}_c &=& \frac{15}{32\pi^2}\int \frac{1}{|\vec{r} - \vec{r}'|} d\vec{r} d\vec{r}' \nonumber \\
            &=& \frac{-15}{64\pi^2} \int |\vec{r} - \vec{r}'| d\vec{S} \cdot d\vec{S}'
\end{eqnarray}
where, the symbols have their usual meanings. Notice that the integrals have
been carried out over nuclear volume, and the lengths have been measured in 
units of the radius parameter $R_o$ of the nucleus with zero deformation. The transformation 
from six dimensional to four dimensional integrals has been accomplished by 
following the technique developed by Kurmanov {\it et al.} \cite{KUR.00}. The 
surface term, on the other hand, is simply modified by the ratio of deformed to 
the corresponding spherical surface areas. The curvature energy term, too, 
needs to be modified to take the deformation effects into account. The modified 
curvature energy ($E_{cur}$) reads:
\begin{eqnarray}
E_{cur}~=~\frac{E_{cur}^{0}}{8\pi} \int_\Omega \left( \frac{1}{R_1} ~+~\frac{1}{R_2}
\right) dS
\end{eqnarray}
where, $E_{cur}^{0}$ is curvature energy at zero deformation; $R_1$ and
$R_2$ are the principal radii of curvature of the nuclear surface 
(in the units of $R_o$), defined by
$r=r_s$; and $dS$ refers to the area element of the nuclear surface. 
The surface parametrisation assumed in the present work is given by:
\begin{eqnarray} 
r_s~=~CR_0(1~+~\sum_{\lambda,\mu}\alpha_{\lambda,\mu}Y_{\lambda,\mu})~.
\label{def}
\end{eqnarray} 
Here, the $Y_{\lambda,\mu}$ functions are the usual spherical harmonics and the
constant $C$ is the volume conservation factor (the volume enclosed by the
deformed surface should be equal to the volume enclosed by an equivalent
spherical surface of radius $R_0$):
\begin{eqnarray} 
C~=~\left[ \frac{1}{4\pi}\int_{\Omega}\left\{1~+~\sum_{\lambda,\mu}
\alpha_{\lambda,\mu}Y_{\lambda,\mu}(\Omega)\right\}^3 d\Omega \right]^{-1/3} .
\end{eqnarray} 
The term $Z^2/A$, which is the correction to Coulomb energy due to surface 
diffuseness of the charge distribution, does not have any explicit deformation 
dependence. This is because the distance function chosen here is 
such that the surface thickness is the same in all the directions 
(see discussion about this in Ref. \cite{BHA.10}).

The total binding energy of a nucleus with $N$ neutrons, $Z$ protons 
and deformation parameters $\beta_2$, $\beta_4$ and $\gamma$ is 
given by:
\begin{eqnarray} 
E\left(N,Z,\beta_2, \beta_4, \gamma\right)
~=~E_{LDM}\left(N,Z,\beta_2, \beta_4, \gamma\right) ~+~
  \eta \, \delta E\left(N,Z,\beta_2, \beta_4, \gamma\right)
\end{eqnarray} 
where, $\delta E$ represents the microscopic part of the binding energy
(shell correction plus pairing energy). The microscopic part has been multiplied
by a factor $\eta$, which is chosen to be 0.85. One of the reasons for introducing 
such a factor is that the Coulomb potential used in the present work 
is less repulsive near $r=0$ than the corresponding value obtained
by using the hard sphere approximation, used in the fit of proton 
mean field (see discussion on this point in Ref. \cite{BHA.10}).  

The free parameters of the liquid drop formula are determined by minimising 
the $\chi^2$ value in comparison with the experimental binding energies
\cite{WAP.03}:
\begin{eqnarray}
\chi^2~=~\frac{1}{n}\sum_{j=1}^{n}
      \left[ \frac{ E(N_j,Z_j) - E_{expt}^{(j)} }
                  { \Delta E_{expt}^{(j)}} \right]^2 ,
\end{eqnarray} 
where $E(N_j,Z_j)$ is the calculated total binding energy for the given
nucleus, $E_{expt}^{(j)}$ is the corresponding experimental value 
\cite{WAP.03}, and $\Delta E_{expt}^{(j)}$ is the uncertainty
in $E_{expt}^{(j)}$.
In the present fit, for simplicity, $\Delta E_{expt}^{(j)}$ 
is set to 1 MeV.

To obtain these parameters 
we proceed as follows. We start by setting in the liquid drop mass formula (\ref{LDM})
the values obtained in our spherical calculation \cite{BHA.10}. Explicitly,
these values are: $a_v$ = -15.841 MeV, $a_s$ = 19.173 MeV, $k_v$ = -1.951,
$k_S$ = -2.577, $r_0$ = 1.187 fm and $C_4$ = 1.247 MeV.
Next, we choose a set of 561 even-even nuclei with $Z\ge 8$ and $N \ge 8$, 
the list of which may be found at \cite{UB.11}.
This set comprises doubly magic, semi magic as well as open shell nuclei,
many of which are expected to be deformed. The main task now is to determine the liquid 
drop parameters as well as the optimal deformation parameters. The calculation proceeds
in the following steps:
\begin{enumerate}
\item Assuming the previously reported \cite{BHA.10} values of the liquid drop 
parameters, the binding energies of these nuclei are obtained by minimising 
on a range of $\beta_2$ values ($\beta_4$ is set to zero in this step). This 
gives a preliminary estimation of $\beta_2$. Next, keeping this $\beta_2$ fixed,
$\beta_4$ is varied to obtain minimum energy. Thus, we now have preliminary 
values of both the deformation parameters.
\item In the next step, keeping the deformation parameters fixed as obtained in 
the earlier step, the liquid drop parameters are fitted by minimising $\chi^2$.
\item With the new values of liquid drop parameters, the deformation parameters
are obtained once again as described in step 1, followed by a final re-fit to the 
liquid drop parameters. 
\end{enumerate}

The numerical values of the new constants of the liquid drop formula obtained 
through this minimisation procedure are: 
$a_v$ = -15.435 MeV, $a_s$ = 16.673 MeV, $a_{cur}$ = 3.161 MeV,
$k_v$ = -1.874, $k_S$ = -2.430, $k_{cur} = 0$ (see discussion below), 
$r_0$ = 1.219 fm, $C_4$ = 0.963 MeV, $w_1$ = -2.762 MeV and $w_2$ = 3.725. 
The values of volume, surface and Coulomb coefficients differ from those 
reported earlier \cite{BHA.10}, primarily due to the inclusion of curvature and 
Wigner terms and the deformation effects. The curvature term, as described
earlier, depends on the mean curvature of the nucleus, which is a function of 
the geometry of the nuclear surface. Therefore, the curvature energy, a priory, 
is expected to modify the surface energy term as well as the $Z^2/A$ term, which
is the correction due to the surface diffuseness of the charge density term.
The somewhat smaller value of the volume coefficient reported here, is not 
surprising. The reduction is due to the influence of the curvature term, as
it has also been found by Pomorski and Dudek (see Table 1 of Ref. \cite{POM.03}).

It is to be noted that the coefficient of the isospin dependent term in the 
curvature energy is very difficult to determine with experimental masses. 
In our case the resulting statistical error in the corresponding 
parameter turns out to be more than 50\% of the numerical value of the 
coefficient. Further, this term is found to weaken the strength of the isospin 
dependent term in the surface energy by a factor of 5. The isospin dependence in
the curvature term, therefore, has been dropped from the present investigation. 

\begin{figure}[htb]
\centerline{\epsfig{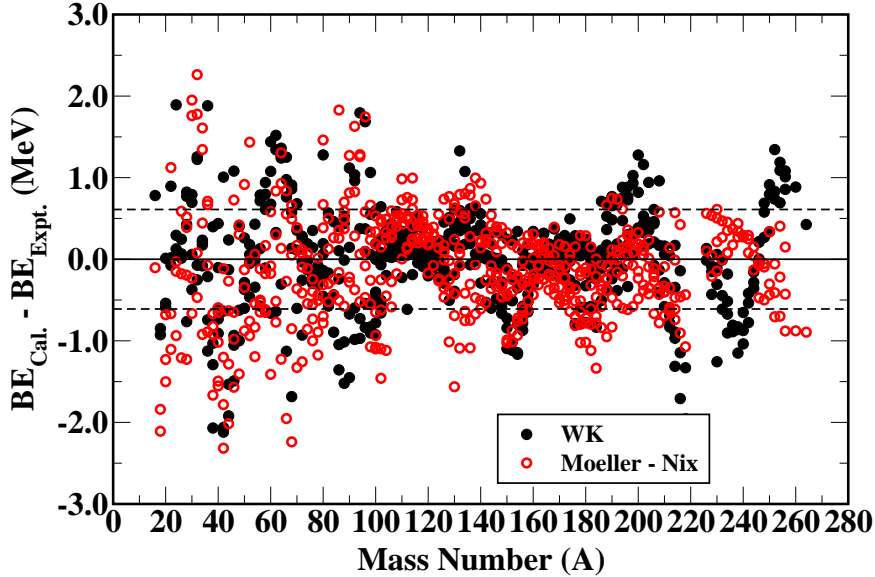}}
\caption{Difference between the calculated (fitted) and the corresponding 
experimental \cite{WAP.03} binding energies, as a function of mass number. 
The dashed horizontal lines correspond to $\delta BE$ = 610 keV. 
The corresponding differences obtained by using the M\"oller - Nix binding 
energies are also presented for comparison.}
\label{dif}
\end{figure}

The {\it rms} deviation of the calculated binding energies with respect to the 
experiment obtained is 610 keV. The M\"oller - Nix calculations \cite{MN.97}, 
for the same set of nuclei, yield a deviation of 656 keV. The explicit values 
of binding energies of our selected set of 561 even-even nuclei used in the 
minimisation procedure can be found at \cite{UB.11}. The present calculation 
establishes that our model is indeed capable of reproducing binding 
energies of deformed nuclei as well, with excellent accuracy.
The difference between the calculated and the corresponding evaluated 
\cite{WAP.03} binding energies is presented in Fig. \ref{dif}. The 
corresponding differences obtained for the M\"oller - Nix calculations is 
presented in the same figure for comparison. The excellent agreement 
found between the calculations and experiment is amply clear from the figure. 

\begin{figure}[htb]
\centerline{\hbox{ \epsfig{file=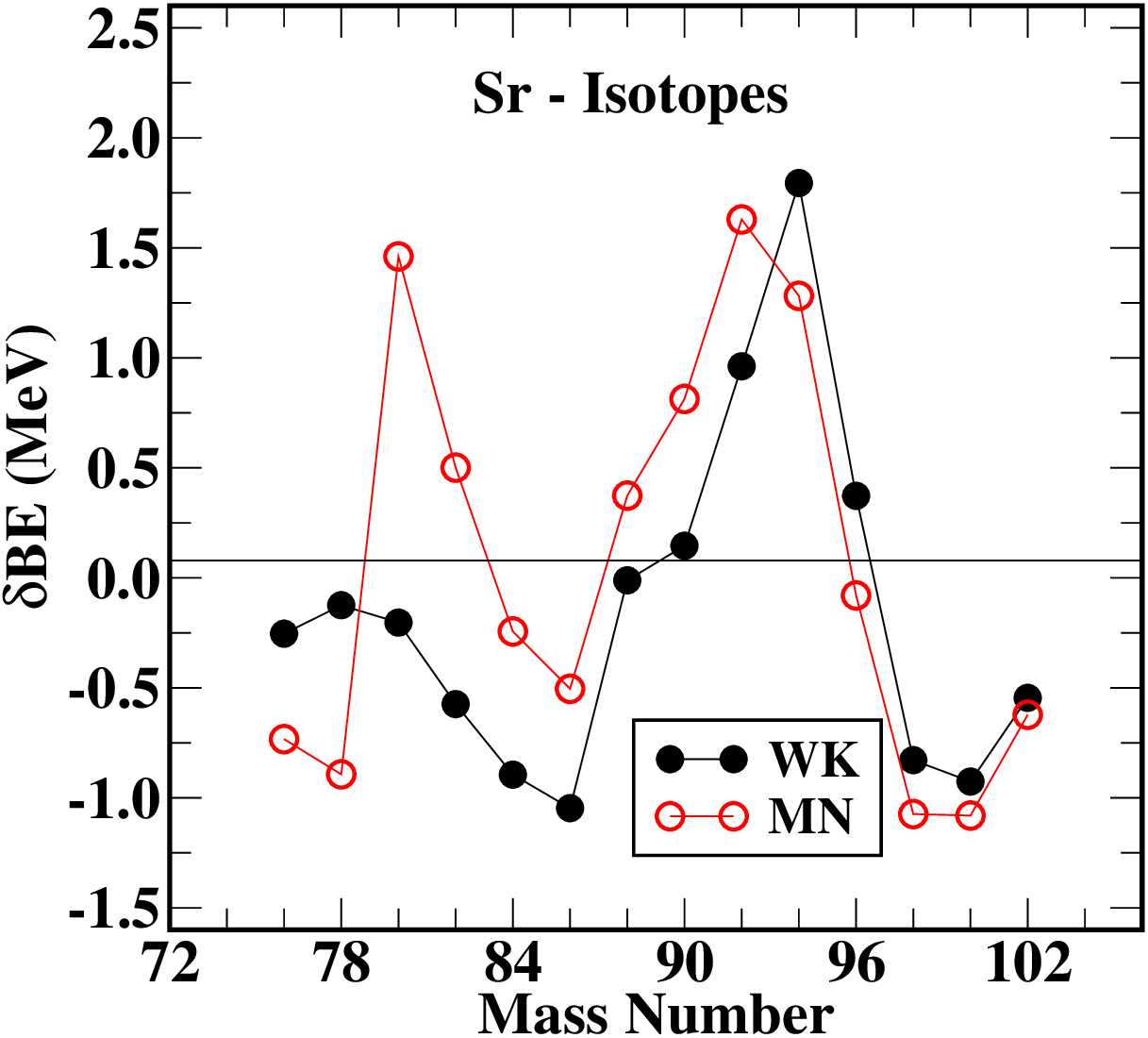,width=0.46\textwidth} \hfill
                   \epsfig{file=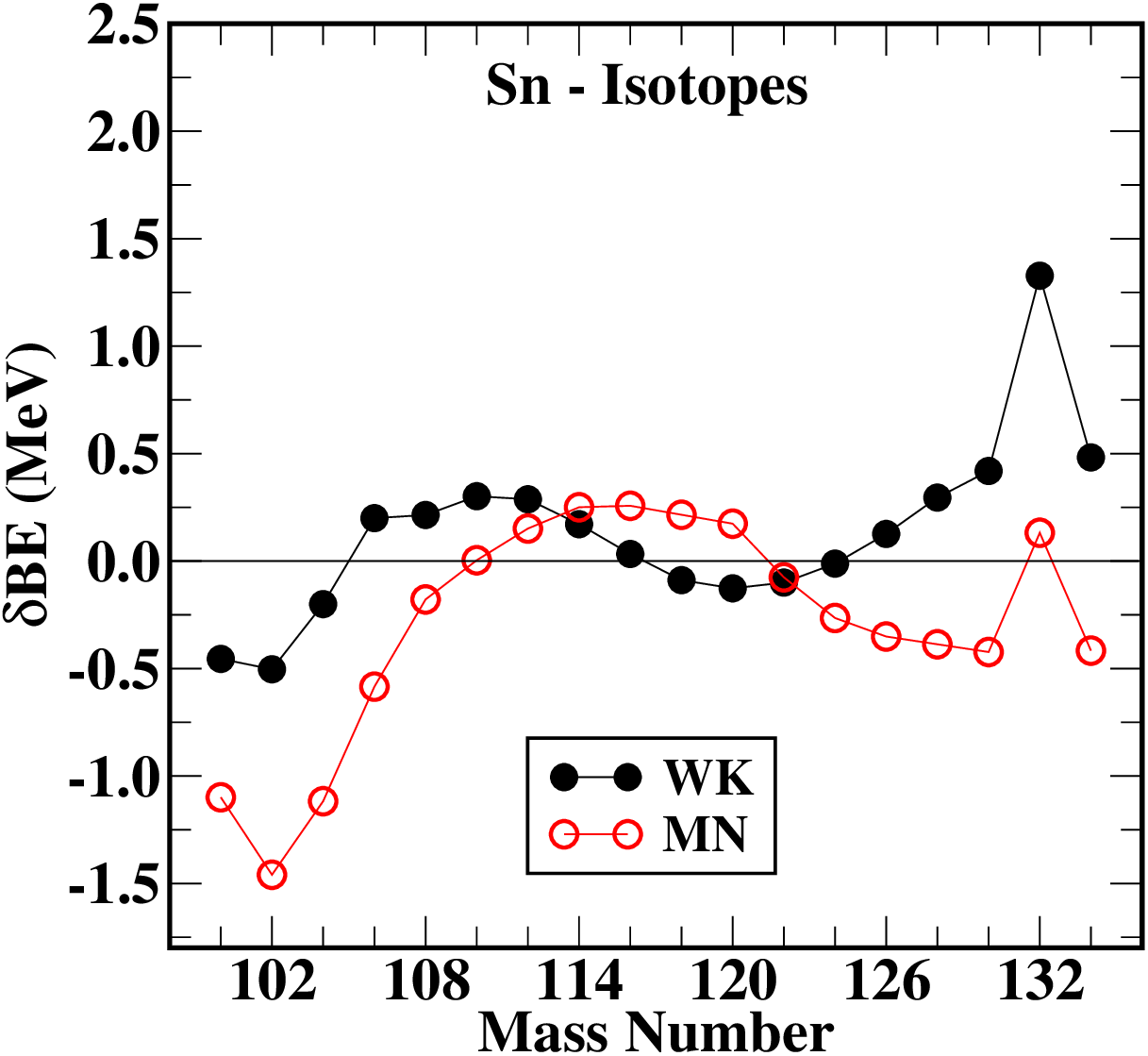,width=0.46\textwidth}}}
\caption{The difference between the calculated and the experimental 
\cite{WAP.03} binding energies for Sr and Sn isotopes.} 
\label{s2n1}
\end{figure}

\begin{figure}[htb]
\centerline{\hbox{ \epsfig{file=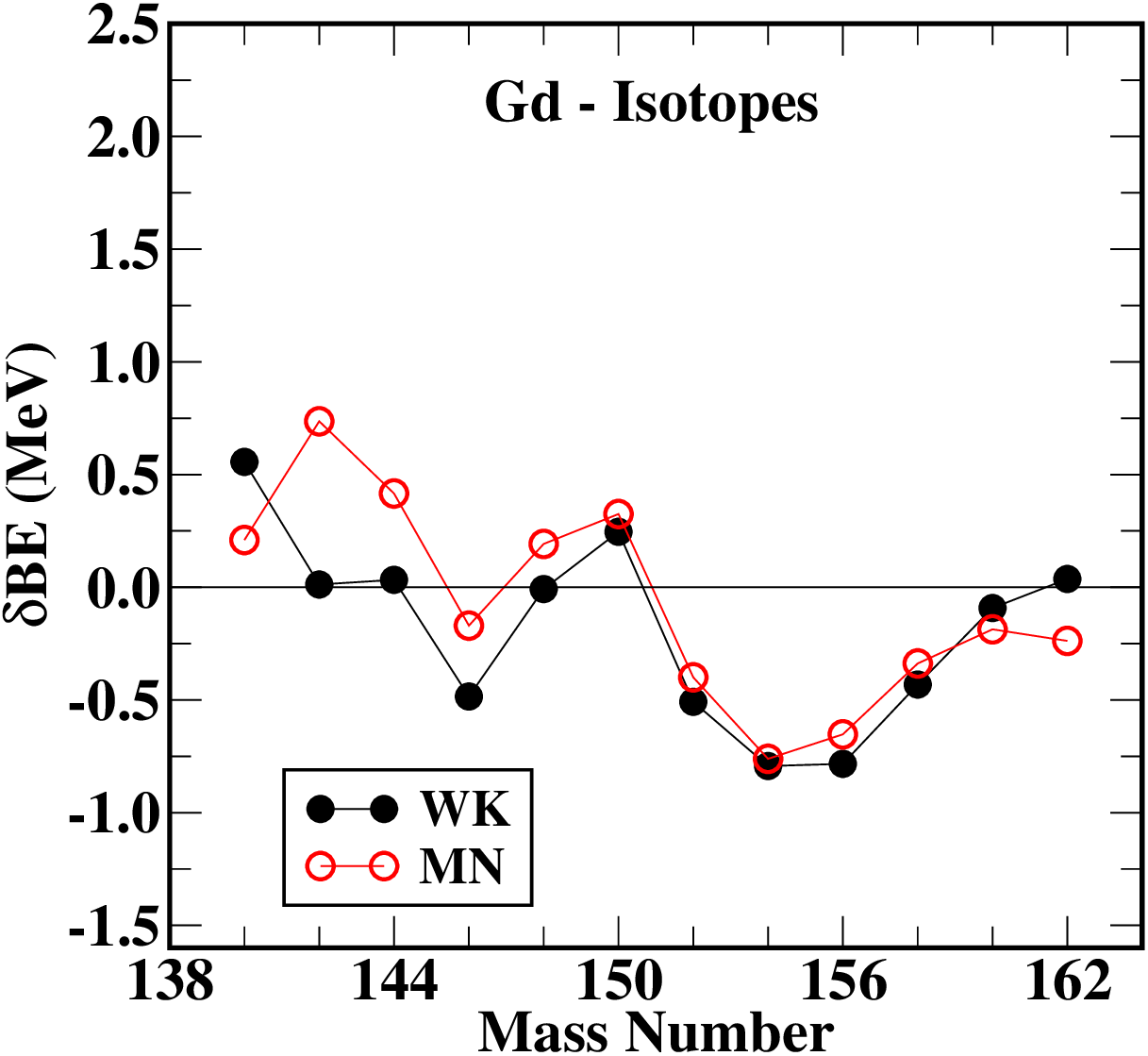,width=0.46\textwidth} \hfill
                   \epsfig{file=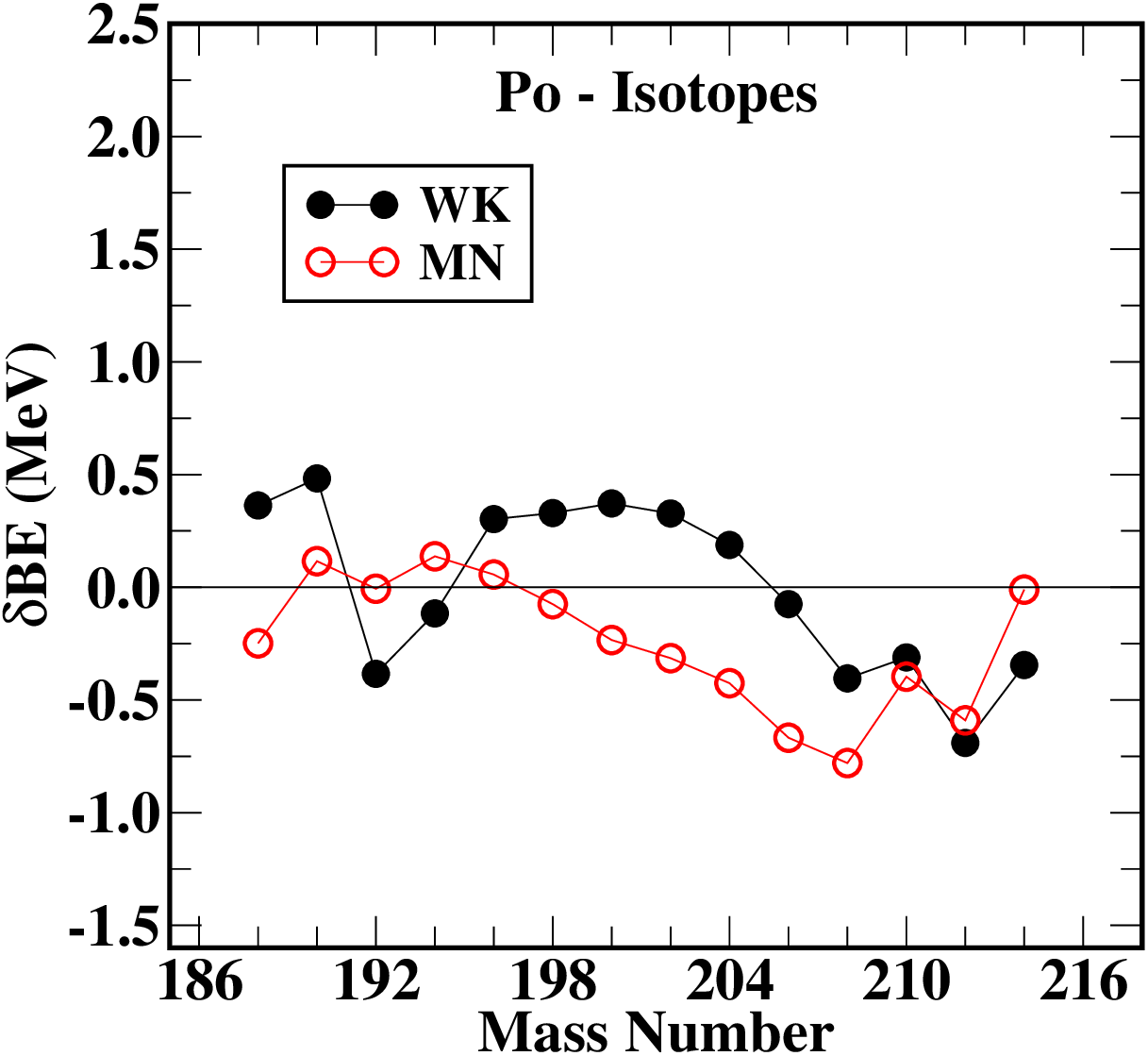,width=0.46\textwidth}}}
\caption{The difference between the calculated and the experimental 
\cite{WAP.03} binding energies for Gd and Po isotopes.} 
\label{s2n2}
\end{figure}

\begin{figure}[htb]
\centerline{\hbox{ \epsfig{file=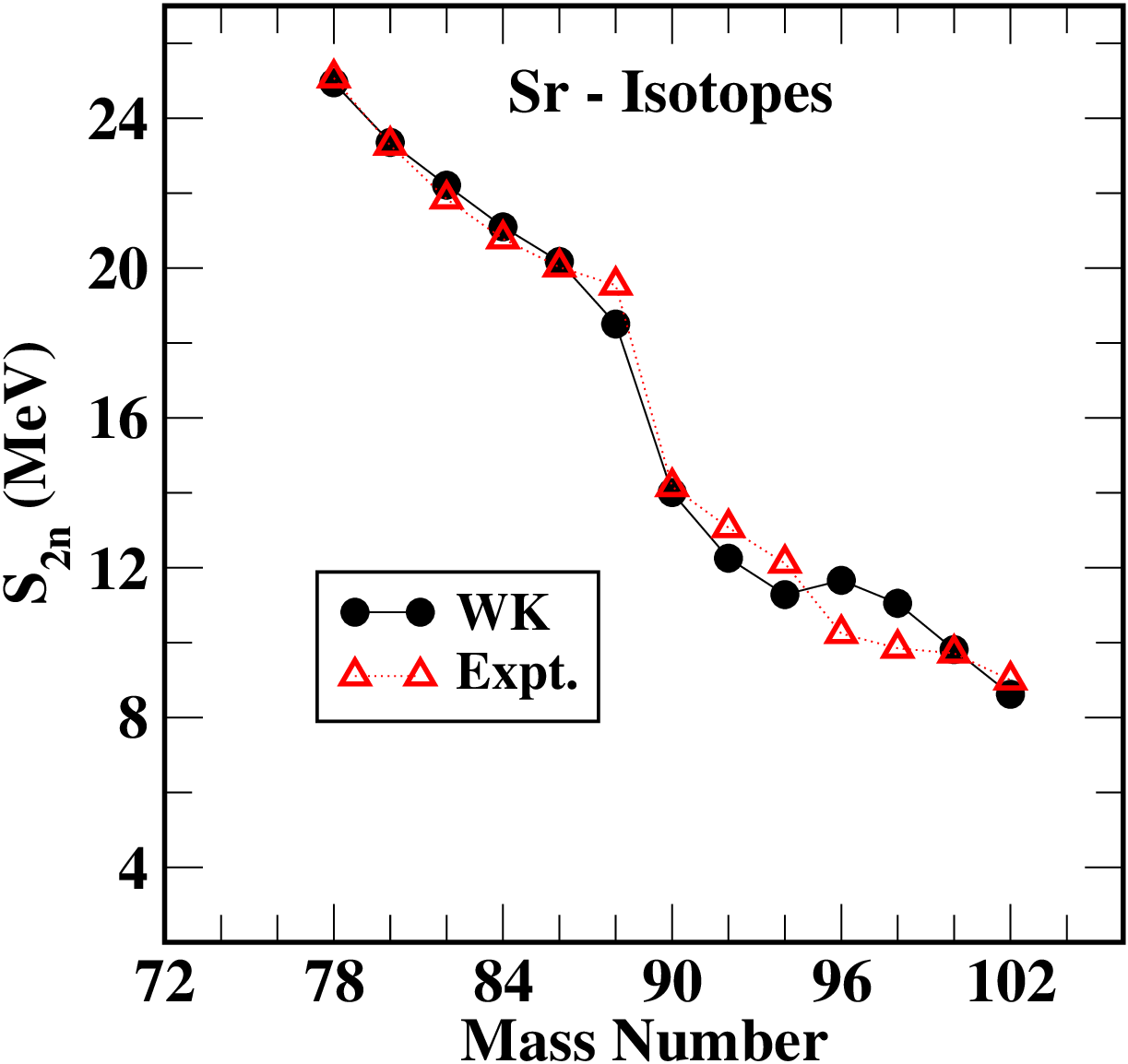,width=0.46\textwidth} \hfill
                   \epsfig{file=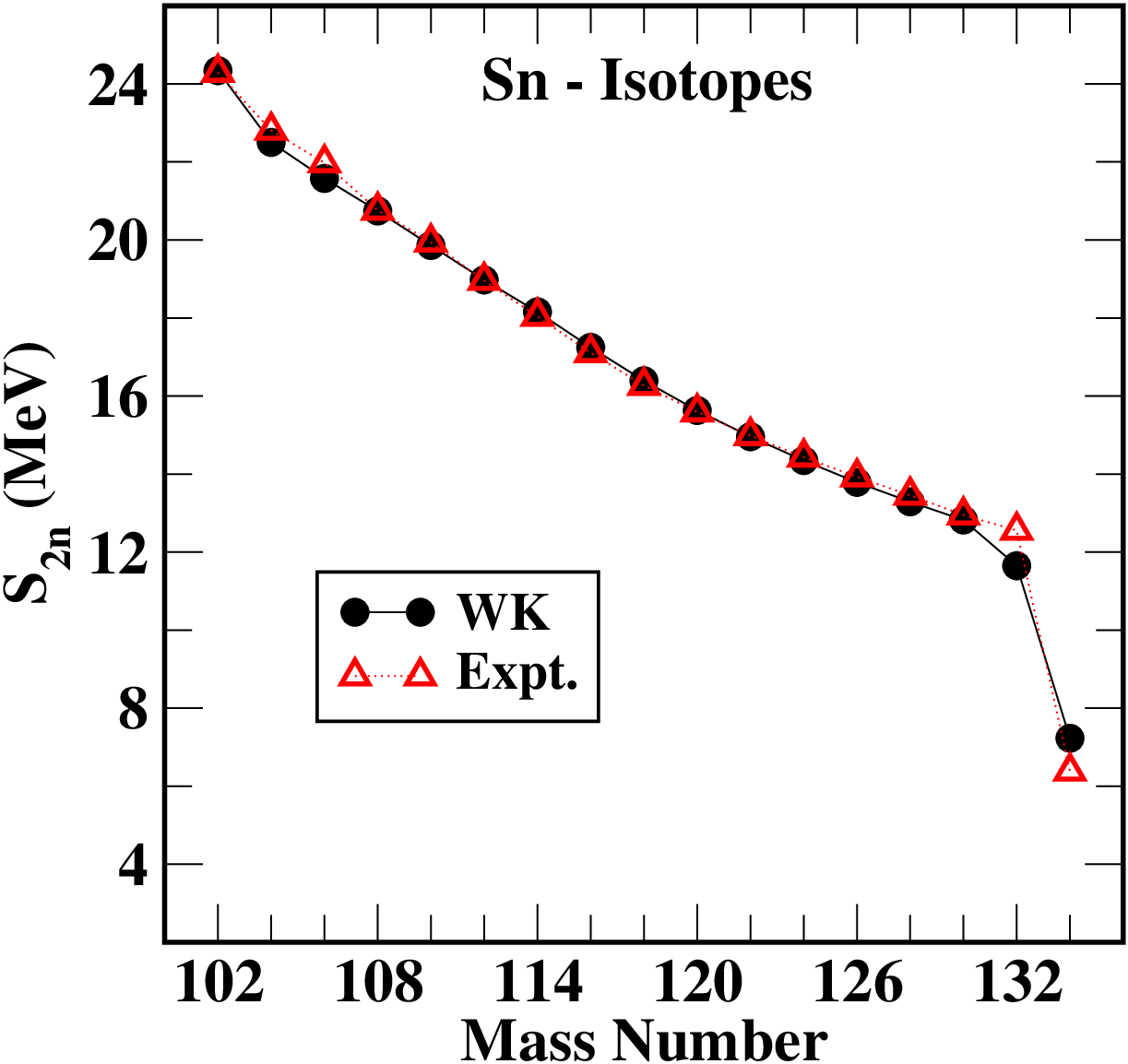,width=0.46\textwidth}}}
\caption{The calculated and the experimental 
\cite{WAP.03} two - neutron separation energies for Sr and Sn isotopes.} 
\label{sr-sn}
\end{figure}

\begin{figure}[htb]
\centerline{\hbox{ \epsfig{file=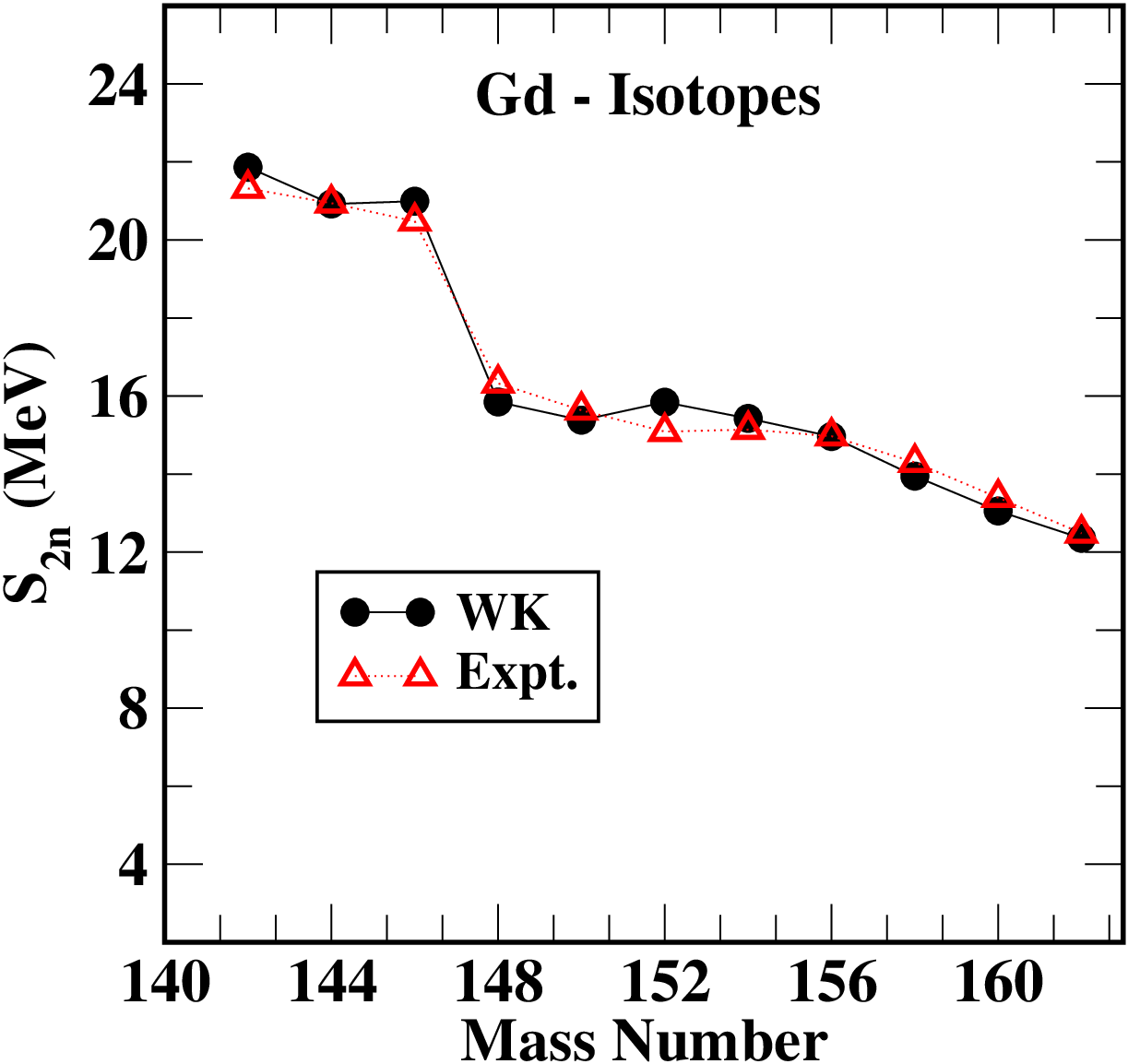,width=0.46\textwidth} \hfill
                   \epsfig{file=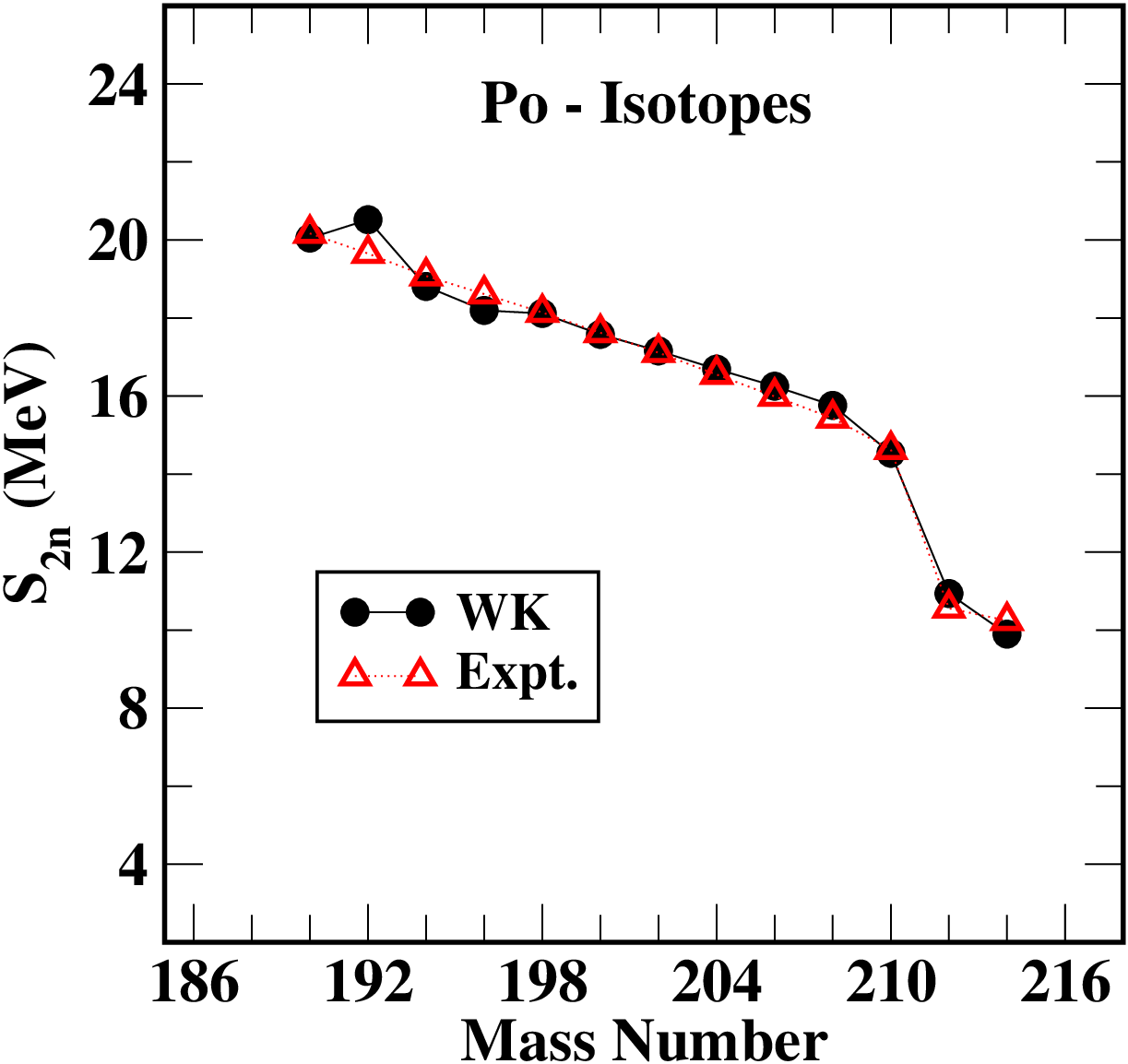,width=0.46\textwidth}}}
\caption{The calculated and the experimental 
\cite{WAP.03} two - neutron separation energies for Gd and Po isotopes.} 
\label{gd-pb}
\end{figure}

We next present and discuss the results obtained for Sr, Sn, Gd and Po 
isotopes as illustrative examples. The difference between the fitted 
and the corresponding experimental binding energies for these nuclei are 
plotted in Figs. \ref{s2n1} and \ref{s2n2}, along with the corresponding 
differences obtained from the M\"oller - Nix calculations \cite{MN.97}. 
The figures reveal that the calculated binding energies (denoted by WK) 
are quite close to the experimental values. The differences are found 
to vary quite smoothly as a function of mass number.
Next, we present the two - neutron separation energies for these chains. The 
two - nucleon separation energies highlight the shell structure in an isotopic
chain. Correct prediction of these separation energies is crucial for 
determination of the drip lines. The calculated and the corresponding 
experimental \cite{WAP.03} two - neutron separation energies are plotted in 
Figs. \ref{sr-sn} and \ref{gd-pb}. The figures reveal that the present 
calculations reproduce the experimental separation energies very well and
that the shell - gaps are also reproduced nicely.

\begin{figure}[htb]
\centerline{\epsfig{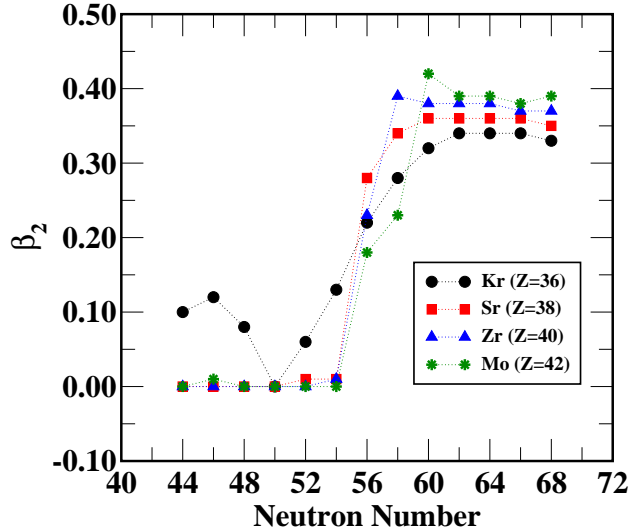}}
\caption{Deformation parameter $\beta_2$ for Kr, Sr, Zr and Mo isotopes.}
\label{defor}
\end{figure}

In addition, the systematics of deformation parameters obtained in these 
calculations turns out to be reasonable. As an illustrative example, we focus 
on the Sr - Zr region. It is well known from the systematics of experimentally 
measured charge radii \cite{ANG.04} that the charge radii increase dramatically 
by 2\% for $^{97}$Rb, $^{98}$Sr and $^{100}$Zr, in comparison to their 
respective lighter isotopes. This jump may be attributed to the possibility 
of onset of highly deformed shapes in the ground - state, around this neutron 
number (see, for example, \cite{HEM.04}).
Our calculations, too, reveal existence of highly deformed ground - states 
(with $\beta_2 \sim 0.3$) around neutron number 60, in the Sr - Zr region. 
The values of $\beta_2$ obtained in this work for Kr, Sr, Zr and Mo chains
are plotted in Fig. \ref{defor}. The sudden change in the ground - state 
deformation around the neutron number 60 is very clear from the figure. 

Further, it is also well known that the ground - states of $^{72}$Kr, $^{76}$Sr 
and $^{80}$Zr have very large ($\sim 0.4$) deformation. This is known to be 
due to population in the intruder $1g_{9/2}$ state. Thus, the ground - state of 
$^{80}$Zr is a 12 particle - 12 hole state, which is manifested again by an 
extremely large stable deformation in the ground - state of $^{80}$Zr. This has 
been verified independently, for example, by the relativistic mean field 
calculation \cite{GAM.92}, density dependent Hartree Fock calculation with 
Skyrme interaction \cite{ZHE.91}, as well as by the Hartree Fock band mixing 
calculation \cite{SAH.90}. The deformation parameters reported in the 
M\"oller - Nix table \cite{MN.95}, too, are consistent with the discussion 
above. It is gratifying to note that the present calculations, indeed, yield 
$\beta_2$ = -0.36, -0.41 and 0.44 respectively, for  $^{72}$Kr, $^{76}$Sr and 
$^{80}$Zr, which is in tune with the mean field as well as the mic - mac 
M\"oller - Nix calculations cited above.

\section{Applications of the Present Model to Near Drip Line Nuclei and Superheavy
Nuclei}
We next test the ability of the present model to describe binding energies 
of the neutron rich and neutron deficient nuclei, as well as of the superheavy
nuclei. To this end, we now present a few exploratory calculations.

\subsection{Proton Drip-line Nuclei in the Ge - Kr Region}

\begin{table}[htb]
\caption{The binding energies and one proton separation energies for 
proton rich nuclei. `Calc.' (MN) represent the results obtained in the 
present work (by M\"oller and Nix \cite{MN.97}). The experimental 
binding energies have been obtained from mass excess values reported
by Tu {\it et al.} \cite{XLT.11}. The experimental $S_p$ values 
have been also been adopted from Ref. \cite{XLT.11}.}
\begin{center}
\begin{tabular}{|c|c|c|c|c|c|c|c|c|} \hline 
           &     &    & \multicolumn{3}{|c|}{Binding Energy (MeV)} & \multicolumn{3}{|c|}{$S_p$ (MeV)}  \\ \cline{4-6} \cline{6-9}
           & $\beta_2$ & $\beta_4$ &  Calc. & MN   & Expt. & Calc. & MN  & Expt. \\ \hline
 ~$^{63}$Ge~ & ~+0.200~   &  ~-0.010~  &~-529.795~   &~-529.266~   &~-530.327~  & ~2.557~  & ~3.315~  & ~2.210~  \\            
 $^{65}$As & +0.210   &  -0.030  &-545.168   &-544.642   &-545.699  & 0.633  & 0.124  & -0.090 \\            
 $^{67}$Se & +0.220   &  -0.050  &-560.598   &-560.158   &-560.698  & 2.379  & 3.364  & 1.852  \\            
 $^{71}$Kr & -0.330   &   0.010  &-592.047   &-591.219   &-591.150  & 2.304  & 3.093  & 2.184  \\ \hline
\end{tabular}
\end{center}
\end{table}
The masses of $^{63}$Ge, $^{65}$As, $^{67}$Se and $^{71}$Kr have recently been 
measured \cite{XLT.11}. These nuclei are very proton rich, and are expected to 
be close to the drip - line. Notice that these nuclei are odd - even and even - 
odd. In this preliminary test of our model near proton drip line, we use the 
simple uniform filling approach for the calculation of the pairing energy. The 
calculated binding energies and one proton separation energies ($S_p$) for 
these nuclei, along with the corresponding experimental values \cite{XLT.11} 
and those reported by M\"oller and Nix \cite{MN.97} are presented in 
Table 1. The binding energies as well as $S_p$ values obtained in the present 
work are found to be quite close to the experiment. This indicates that the 
present model extrapolates reliably up to the proton drip lines. The nucleus 
$^{65}$As is reported to be slightly unbound against proton emission with $S_p = -90 \pm 85$ keV 
\cite{XLT.11}. Our calculation, on the other hand, yields a positive value of 
$S_p$ for $^{65}$As, indicating a proton bound nucleus. However, it should be 
noted that the separation energies are obtained by taking differences of the 
relevant binding energies, and hence are very sensitive to the precise details 
of the same. The fact that the theoretical separation energies obtained in 
this work differ from the corresponding experimental values only by a few
hundred keV's is quite remarkable.

\subsection{Composition of the Outer Crust of Neutron Stars}
The masses of very neutron-rich nuclei are particularly interesting
for some astrophysical calculations. We next compute the composition
of the outer crust of a neutron star as a further application of our
present mass model. As one moves from the surface of a neutron star to
its interior, the outer crust is the region comprising matter at
densities between $\sim\! 10^{4}$ g/cm$^3$ and $\sim\! 10^{11}$
g/cm$^3$. Matter at those densities consists of fully-ionised,
neutron-rich atomic nuclei that arrange themselves in the lattice
sites of a Coulomb crystal embedded in a degenerate electron gas \cite{HAE.07}. The
neutron excess of the nuclei in the outer crust becomes
larger with increasing matter density until neutron drip starts taking
place at a density about $4\times 10^{11}$ g/cm$^3$. At that point,
one leaves the outer crust and enters the so-called inner crust of the
neutron star, where the atomic nuclei are immersed in an electron gas
and a neutron gas.

In order to compute the composition of the outer crust we follow the
usual formalism as described in Refs.~\cite{RUSTER.06,ROCA.08,ROCA.12}
and references quoted therein. That is, we consider cold and
electrically neutral matter which is assumed to be in thermodynamic
equilibrium and in its absolute ground - state. We calculate the Gibbs
free energy of this system by adding the contributions of the nuclear,
electronic, and lattice terms \cite{RUSTER.06,ROCA.08,ROCA.12} and,
finally, we evaluate the equilibrium composition ($Z$,$N$) at a
certain pressure by minimising the obtained Gibbs free energy per nucleon. 

\begin{figure}[t]
\centerline{\epsfig{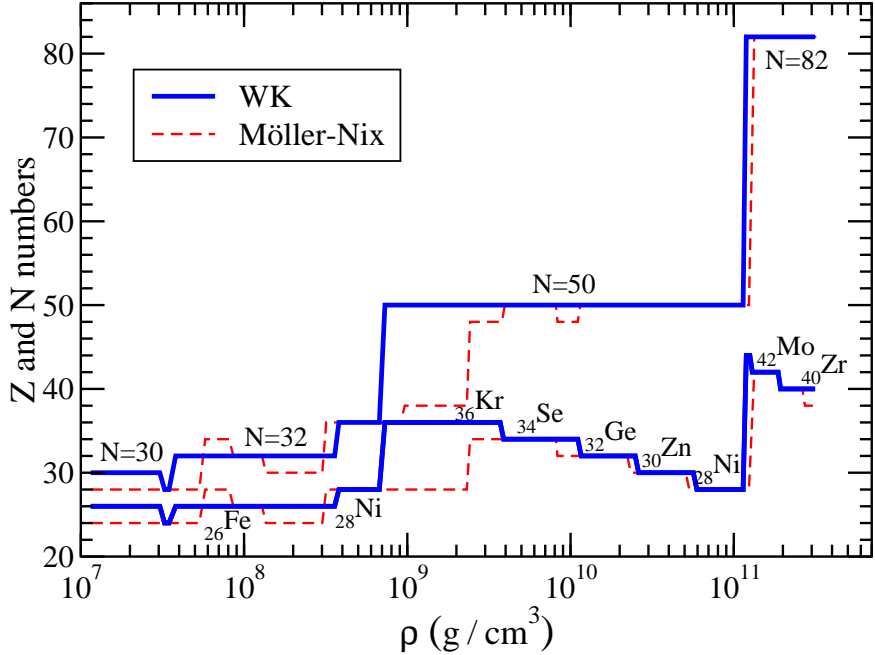}}
\caption{Predicted composition of the outer crust of a neutron star as
a function of the density. The upper line depicts the variation of the
neutron number $N$, while the lower line depicts the variation of the
proton number $Z$. The composition obtained by using the M\"oller-Nix
mass formula is also presented for comparison.}
\label{crust}
\end{figure}

We display our predictions for the equilibrium nuclear species present
in the outer crust in Fig.~\ref{crust}. We perform the calculations within
the range $\rho = 10^{7}$ g/cm$^3$ to $\rho = 3 \times 10^{11}$ g/cm$^3$. 
The variation of the neutron
and proton numbers with increasing crustal density shows a structure
of plateaus that are interrupted by abrupt jumps in the composition.
As exemplified by the $N=50$ plateau, the prevalence of a given
nucleon number over a large range of densities is related with the
shell effect due to the filling of a nuclear shell. The $N=50$ neutron
plateau also is very illustrative of the fact that, with increasing
density, it is energetically favorable for the nuclei of the crust to
capture electrons from the degenerate electron gas. This results in
increasingly neutron-rich nuclides along the neutron plateau.
Eventually, the mismatch between the neutron and proton numbers is too
large and the jump to the next neutron plateau takes place in an
effort to reduce the penalty imposed on the system by the nuclear
symmetry energy \cite{ROCA.08,ROCA.12}.

At low crustal densities up to about $\rho= 7 \times 10^{8}$ g/cm$^3$,
our calculations sequentially favor $^{56}_{26}$Fe,
$^{58}_{26}$Fe, and $^{64}_{28}$Ni as the equilibrium nuclides (with
$^{52}_{24}$Cr occurring in a short density interval between
$^{56}_{26}$Fe and $^{58}_{26}$Fe). Once the jump to the $N=50$ plateau
ensues at a density $\rho \sim\! 7 \times 10^{8}$ g/cm$^3$, our
model predicts the sequence of increasingly neutron-rich isotones
$^{86}_{36}$Kr, $^{84}_{34}$Se, $^{82}_{32}$Ge, $^{80}_{30}$Zn, and
$^{78}_{28}$Ni. After the $^{78}_{28}$Ni nucleus, it is unfavorable to
move further to $^{76}_{26}$Fe and at a density $\rho \sim\! 1.2 \times
10^{11}$ g/cm$^3$ we find that the composition of the crust jumps
to the $N=82$ plateau (where our calculations predict the occurrence of
the isotones $^{124}_{42}$Mo and $^{122}_{40}$Zr). 
We display the results obtained with the M\"oller-Nix mass table
\cite{MN.95} in the same Fig.~\ref{crust} for comparison. Though the
overall pattern is quite similar to the results obtained with our
calculated masses, the M\"oller-Nix mass table predicts more structure
in the variation of the neutron and proton numbers with the crustal
density, and the jump to the $N=50$ plateau is delayed to a little
higher density. This fact suggests that in the present mass region the
shell effects due to the filling of nuclear shells and sub-shells are
somewhat weaker in the M\"oller-Nix mass formula than in our model.

\subsection{Superheavy Nuclei}
Production and study of superheavy nuclei is of current interest from both
theoretical \cite{BEN.00,GOR.02,MUN.03,YKG.05} and experimental 
\cite{OGA.07,HOF.00} aspects. 
With the advent of increasingly sensitive detection methods, it 
is possible to identify the superheavy elements, and measure 
$\alpha$ decay $Q$ values precisely. The elements with $Z=118$ have
been produced so far \cite{OGA.07}. Here, we apply our mic-mac 
model to a few recently reported superheavy nuclei \cite{OGA.07}.  
In particular, we focus on the $\alpha$ decay $Q$ values ($Q_\alpha$).
The binding energies of the parent as well as the daughter nuclei, 
necessary to obtain the $Q_\alpha$ values, are obtained within our 
mic-mac model by minimising over the deformation ($\beta_2$, $\beta_4$) mesh. 
The binding energy of the $\alpha$ particle is adopted from the Audi - Wapstra 
compilation \cite{WAP.03}. The calculated (Calc.) as well as the 
experimental $Q$ values \cite{OGA.07} are presented in Table 2. 
We find that the calculated $Q_\alpha$ values are very close to 
the experiment. This is quite encouraging, since as in the case of 
the separation energies, the $Q$ values as well are obtained by 
taking differences between two large quantities. 

\begin{table}[htb]
\caption{The $\alpha$ decay $Q$ values and half lives ($T_{1/2}$) 
for some of the superheavy nuclei.}
\begin{center}
\begin{tabular}{|c|c|c|c|c|c|} \hline
      &       & \multicolumn{2}{|c|}{$Q_\alpha$ (MeV)} & \multicolumn{2}{|c|}{$T_{1/2}$} \\ \cline{3-4}\cline{5-6}
 $Z$  &  $A$ &  Calc.  &   Expt.         &   Calc. &  Expt. \\\hline
~118~ &~294~ & ~11.76~ & ~11.81$\pm$0.06~&~0.56 ms~&~0.89$^{+1.07}_{-0.31}$ ms~\\
 116  & 293  &  10.59  &  10.69$\pm$0.06 & 136 ms  & 61$^{+57}_{-20}$ ms\\
 116  & 292  &  10.66  &  10.80$\pm$0.07 &  89 ms  & 18$^{+16}_{-6}$ ms \\ 
 116  & 291  &  10.89  &  10.89$\pm$0.07 &  22 ms  & 18$^{+22}_{-6}$ ms  \\ 
 115  & 288  &  10.49  &  10.61$\pm$0.06 & 129 ms  & 87$^{+105}_{-30}$ ms \\
 115  & 287  &  11.38  &  10.74$\pm$0.09 &0.69 ms  & 32$^{+155}_{-14}$ ms  \\
 114  & 289  &   9.91  &   9.96$\pm$0.05 &  2.7 s  & 2.6$^{+1.2}_{-0.7}$ s  \\
 114  & 288  &  10.26  &  10.08$\pm$0.06 & 0.28 s  & 0.80$^{+0.27}_{-0.16}$ s\\
 114  & 287  &  10.19  &  10.16$\pm$0.06 & 0.43 s  & 0.48$^{+0.16}_{-0.09}$ s \\
 113  & 283  &  10.82  &  10.26$\pm$0.09 & 4.6 ms  & 100$^{+490}_{-45}$ ms\\
 113  & 282  &  10.99  &  10.78$\pm$0.08 &  17 ms  & 73$^{+134}_{-29}$ ms  \\
 111  & 280  &   9.33  &   9.87$\pm$0.06 &  17 s   & 3.6$^{+4.3}_{-1.3}$ s  \\
 111  & 279  &  10.56  &  10.52$\pm$0.16 &  5.5 ms  & 170$^{+810}_{-80}$ ms  \\ \hline
\end{tabular}
\end{center}
\end{table}

The $\alpha$ decay $Q$ values can be related to the half lives through
the Viola - Seaborg relation \cite{VIO.66}. In particular, following 
Oganessian \cite{OGA.07}, we adopt:
\begin{eqnarray} 
\log T_{1/2}~=~\frac{aZ~+~b}{\sqrt{Q_\alpha}} ~+~ cZ ~+~ d
\end{eqnarray} 
where, $Z$ is the charge number of the parent nucleus; $Q_\alpha$ is 
the $\alpha$ decay $Q$ value, and $a$, $b$, $c$ and $d$ are parameters,
taken to be \cite{OGA.07}: $a$ = 1.787, $b$ = -21.40, $c$ = -0.2549 
and $d$ = -28.42. The half lives obtained by using 
the calculated $Q$ values are found to be in reasonable agreement 
with the experiment. At places, the calculations do deviate by an order 
of magnitude, but notice that the half lives have very large uncertainties.

\section{Systematic investigation of the factor $\alpha$}

Large scale calculations using the proposed mic-mac model can be cumbersome
and highly time consuming. Therefore, it may be very useful to look for 
simplifications that allow to speed up the calculations without loss of 
accuracy. To this end, we explore the possibility of absorbing the fourth -
order correction 
\begin{eqnarray} 
E_{4} = E_{\hbar^{4}}^{CN}(n) + E_{\hbar^{4}}^{SO}(n) + E_{\hbar^{4}}^{CN}(p) + E_{\hbar^{4}}^{SO}(p)
\end{eqnarray} 
into the net second - order contribution:
\begin{eqnarray}
E_{2} = E_{\hbar^{2}}^{CN}(n) + E_{\hbar^{2}}^{SO}(n) + E_{\hbar^{2}}^{CN}(p) + E_{\hbar^{2}}^{SO}(p)~.
\end{eqnarray} 
Here, $(n)$ and $(p)$ stand for neutronic and protonic contributions. See Eqs. 
(11)-(14) for the definitions of the different terms appearing in these two 
equations. This absorption is expected to have two major effects. 
Clearly, if such an absorption is possible, the factor $\alpha$ 
(see Eqs. (1) and (2) for definition), should be expressible as a function of 
neutron number, proton number, or some combinations thereof.
Before discussing the possibility of absorbing 
fourth - order terms into second - order terms for a Woods - Saxon potential, 
we demonstrate the existence of such a functional form for the simple Harmonic 
Oscillator potential.

\subsection{The Harmonic Oscillator Potential}
The harmonic oscillator (HO) potential provides a unique opportunity to 
investigate the details of the WK expansions analytically. Therefore, first 
we consider the simplest form of the HO potential, without spin - orbit 
interaction. It can be shown that for the HO potential, the different WK 
corrections are given by \cite{JEN.75}:
\begin{eqnarray}
E_{4} &=& \frac{-17\hbar \omega}{960} \\
E_{2} &=& \frac{\lambda^2}{8\hbar \omega}
\end{eqnarray} 
where, $\lambda$ is the chemical potential, determined as described earlier,
and $\omega$ is the oscillator frequency.
For the HO potential, assuming degeneracy of 2, the particle number 
(see Eq. (7)) is given by:
\begin{eqnarray} 
{\cal {N}}~=~ \frac{1}{3} \left( \frac{\lambda}{\hbar\omega} \right)^3 - 
              \frac{1}{4} \left( \frac{\lambda}{\hbar\omega} \right)
\end{eqnarray} 
This equation is cubic in $\lambda/(\hbar\omega)$, and in principle can be 
solved exactly. Here, however, we take an alternative and physically more 
transparent approach, wherein, we express $\lambda$ as \cite{CEN.07,CEN.98}
\begin{eqnarray} 
\lambda~=~\lambda_0 + \lambda_2 + \lambda_4
\end{eqnarray} 
where, $\lambda_j$ is correct up to order $\hbar^j$. Starting from the Thomas 
Fermi expression for the chemical potential, and noticing that the normalisation
is true order by order, we get the following expression for chemical potential,
correct up to $\hbar^4$:
\begin{eqnarray} 
\lambda~=~\left\{ \left ( 3{\cal {N}} \right)^{1/3} + \frac{1}{4} \left ( 3{\cal {N}} \right)^{-1/3} \right\}\hbar\omega
\label{HOLAM}
\end{eqnarray} 
This, along with the second and fourth - order WK corrections to energy 
(see Eqs. (11),(12)), yields
\begin{eqnarray} 
\alpha~=~1~-~\frac{17}{60}\frac{(\hbar \omega)^2}{\lambda_{p}^{2} + \lambda_{n}^{2}}
\label{ALHO}
\end{eqnarray} 
where, $\lambda_{p}$ and $\lambda_{n}$ are chemical potentials for $Z$ protons 
and $N$ neutrons respectively. Further, notice that the neutron and proton 
numbers can be written as:
\begin{eqnarray} 
N~=~\frac{1 + I}{2} A~~~\mathrm{and}~~~Z~=~\frac{1 - I}{2} A
\end{eqnarray} 
$A=N+Z$ being the mass number of the nucleus, and $I$ being asymmetry 
parameter, defined as $I=(N-Z)/A$. We obtain,
\begin{eqnarray} 
\alpha~=~1 - \frac{17}{120} \left( \frac{2}{3} \right)^{2/3} A^{-2/3} \left(1 + \frac{1}{9}I^2 \right)
\end{eqnarray} 
where, the terms up to the order $A^{-2/3}$ are retained, and the expansion in 
$I$ has been carried out only up to second - order in $I$. It can be therefore 
seen that the factor $\alpha$ can indeed be written as a function of mass number
and $I$, implying that it is in principle possible, at least in the case of HO 
potential, to absorb the fourth - order WK corrections to the energy
into the second - order WK corrections.

\begin{figure}[htb]
\centerline{\epsfig{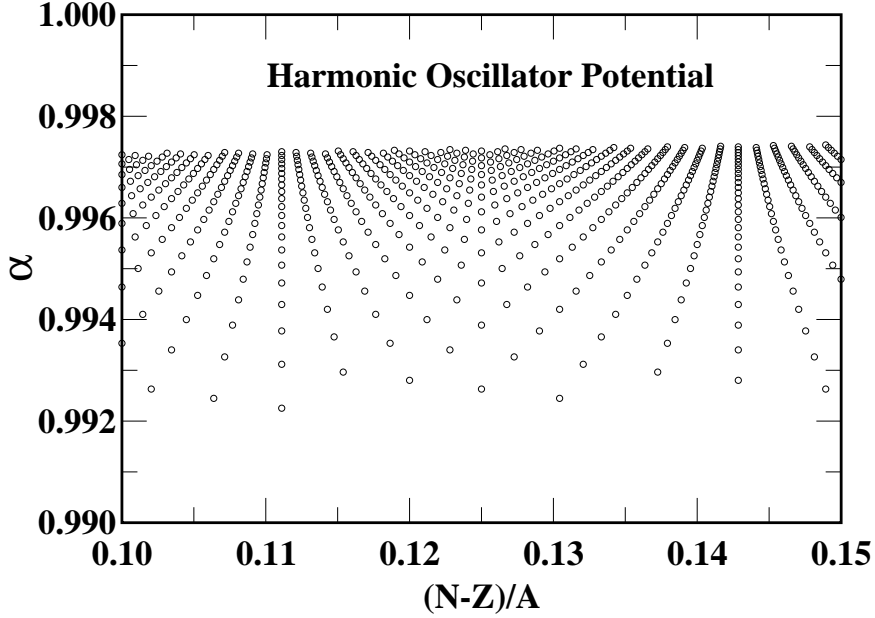}}
\caption{The factor $\alpha$ for Harmonic Oscillator potential, without spin - 
orbit interaction. Only a small portion of the asymmetry scale has been 
presented.}
\label{HOALPF}
\end{figure}

To understand the behaviour of $\alpha$ with respect to $I$, we plot the factor
$\alpha$ as a function of $I$ in Fig. \ref{HOALPF}.
It is seen that the factor $\alpha$ has a very regular behaviour with respect 
to asymmetry. There are points stacked at a given value of $I$, with groups of 
points placed symmetrically with respect to them. This regularity persists over 
the entire range of $I$ values.

\subsection{Woods - Saxon potential}
\begin{figure}[htb]
\centerline{\epsfig{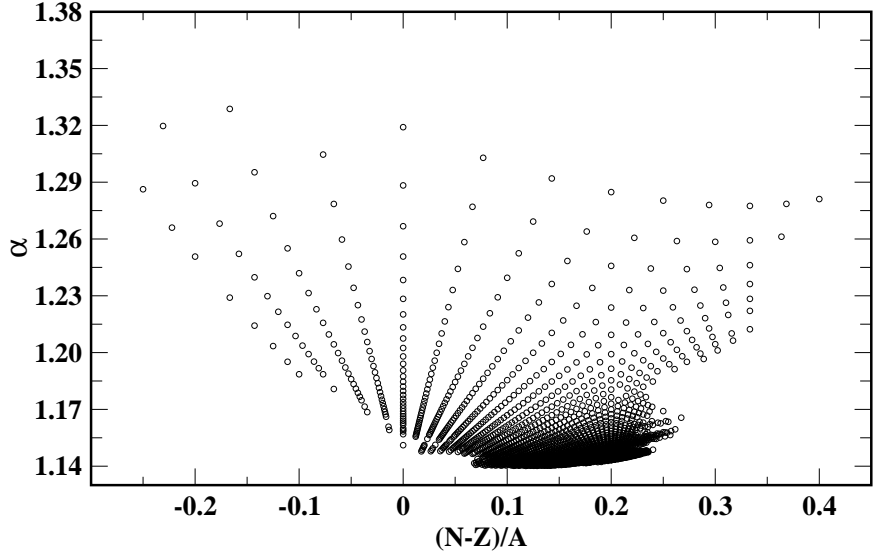}}
\caption{Factor $\alpha$ as a function of asymmetry for a Woods-Saxon 
potential.}
\label{INT_full}
\end{figure}

\begin{figure}[htb]
\centerline{\epsfig{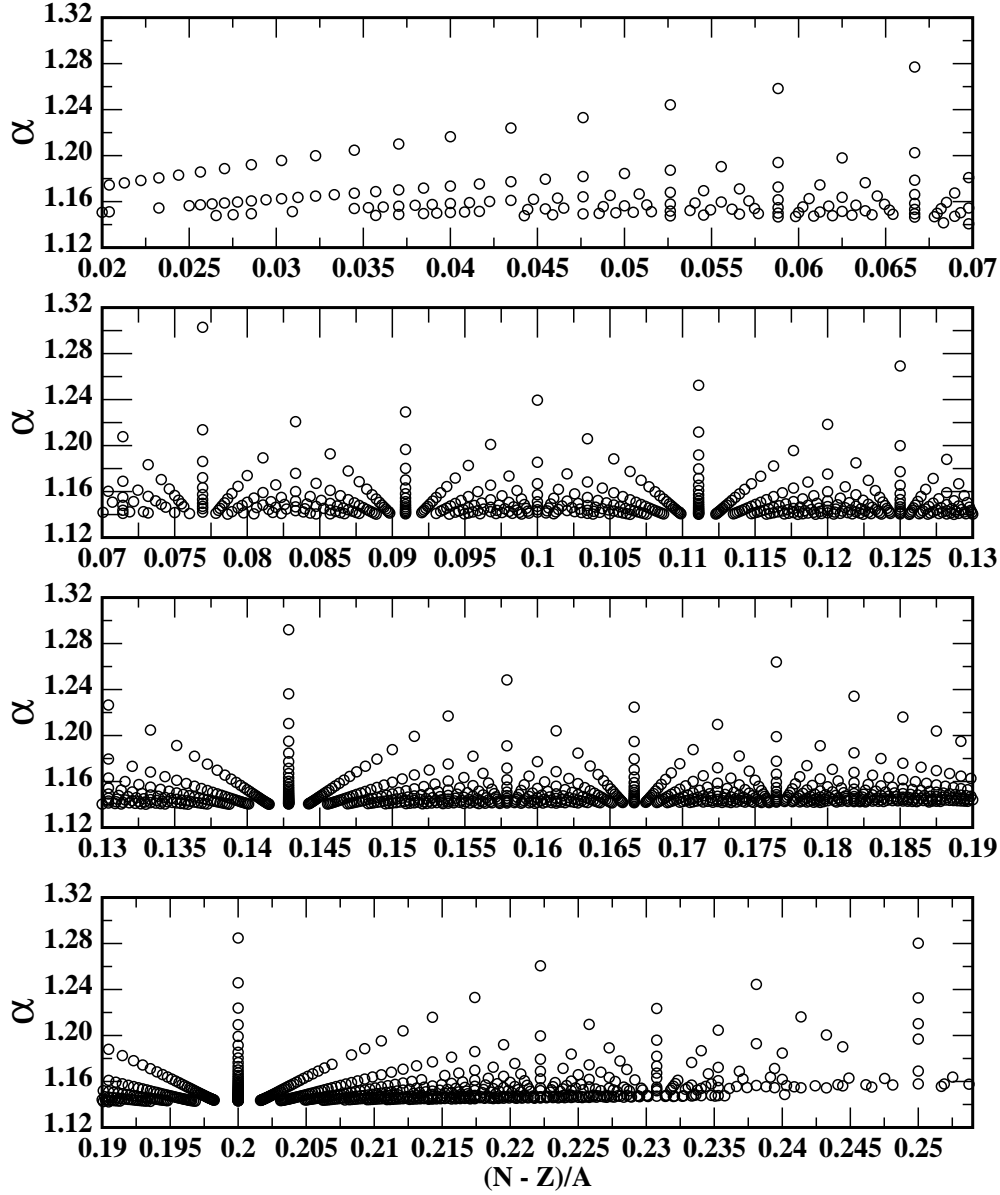}}
\caption{Fate $\alpha$ as a function of asymmetry.}
\label{INT_detail}
\end{figure}

Next, we investigate the factor $\alpha$ for the Woods - Saxon potential.
In order to achieve this, we choose a set of 2171 known nuclei 
\cite{WAP.03} with $Z>5$. Spherical symmetry is 
assumed. The nuclear, spin - orbit and Coulomb potentials have been 
taken as defined in Ref. \cite{BHA.10}. The full Wigner - Kirkwood calculations 
up to the fourth - order in $\hbar$ are carried out for 
these nuclei, and the exact values of the factor $\alpha$ are obtained. 
These are then plotted as a function of the asymmetry parameter $I$ in Fig.
\ref{INT_full}. The figure exhibits that the factor $\alpha$ has a very
regular behaviour as a function of asymmetry. In order to understand the 
detailed structure of the factor $\alpha$, we plot the same results
with a greater resolution in Fig. \ref{INT_detail}. 

A remarkable and regular pattern emerges from the plots. In comparison with the
case of the HO potential, the pattern is inverted. The pattern consists of 
`fan like' structures. There are groups of points stacked exactly along vertical
lines, as indicated in Fig. \ref{INT_detail} accompanied by symmetrically 
placed, slanting groups of points. All these groups of points constitute nearly 
perfect straight lines. This is in contrast with the case of HO potential, where
the lines were curved.

A closer examination of the behaviour of the factor $\alpha$ reveals several 
interesting features. To understand them better, we shall first enlist the 
nuclei appearing in a particular `fan' structure. We 
shall designate the slanting lines appearing in the `fan' structure as `rays'. Thus, 
each fan structure has a number of rays in it, symmetrically placed with respect
to the vertical line, defined by a particular ratio, $(N - Z)/A$. For example, 
let us consider $(N - Z)/A = 1/11$.  This fan structure has $^{22}$Ne, $^{33}$P,
$^{44}$Ca, ..., $^{176}$Hg, ... along the vertical line. The first ray to the 
right of this line contains nuclei like $^{20}$F, $^{31}$Si, $^{42}$K 
$\cdot\cdot\cdot$. The second ray to the right of the vertical line consists of 
the nuclei like $^{40}$Ar, $^{51}$V, $^{62}$Ni etc. The first ray to the left 
of the vertical line consists of $^{35}$S, $^{46}$Sc, $^{57}$Fe etc. Whereas, 
the second ray to the left of the vertical line consists of $^{37}$Cl, 
$^{48}$Ti, $^{59}$Co etc. The heavier nuclei in this sequence are towards the 
bottom of the pattern. The value of $\alpha$ is therefore, inversely 
proportional to the mass number. Thus, it is expected that in the limit of 
$A \rightarrow \infty$, the $\alpha$ values will approach some constant value,
say, $\alpha_0$, which is approximately 1.125, according to the figure above. 

\begin{figure}[htb]
\centerline{\epsfig{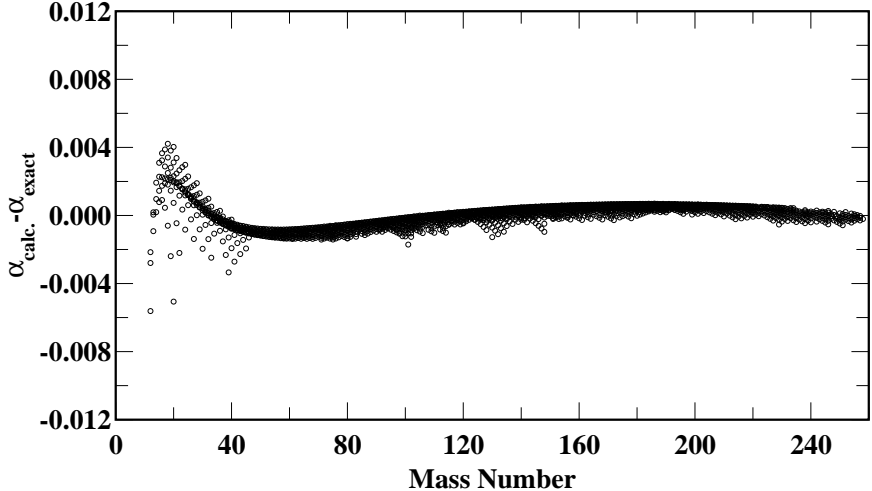}}
\caption{The difference between the fitted and the corresponding 
exact values of $\alpha$.}
\label{New_alpha}
\end{figure}

Considering these observations, we propose the following parametrisation for the factor 
$\alpha$:
\begin{eqnarray}
\alpha = \alpha_0 + \frac{\alpha_1}{A} + \alpha_2\frac{N-Z}{A} 
        + \alpha_3 \left(\frac{N-Z}{A}\right)^2
\label{alpPH}
\end{eqnarray}
where, $\alpha_j$'s are adjustable parameters. Considering all the 2171 nuclei 
(see above), we carry out a least squares fit to determine these parameters. The 
fit turns out to be exceptionally good, with {\it rms} deviation 
$1.09\times 10^{-3}$. The values of the parameters are:
$\alpha_0$=1.12761; $\alpha_1$=2.26744; $\alpha_2$=-0.02659 and 
$\alpha_3$=0.29987. 
The difference between the exact and the corresponding fitted $\alpha$ values is 
plotted in Fig. \ref{New_alpha}, indicating that the agreement is almost 
perfect, and that the phenomenological formula that has been proposed here is 
indeed robust, for all the mass regions.

We shall now investigate the deformation effects particularly with reference to 
the factor $\alpha$. In order to achieve that, we once again consider the set 
of 561 even - even nuclei (see Section 3), with deformation parameters obtained 
as described before.
The calculation of binding energies requires the shell corrections, pairing 
energies and the liquid drop energies. The shell corrections require averaged 
energies, which are calculated here using the WK expansion. Here, we consider 
the WK expansion only up to second - order, and simulate the effects of fourth -
order through the factor $\alpha$ (Eq. (\ref{alpPH})). This defines the averaged
energies and hence the shell corrections completely. The difference between the 
shell corrections thus obtained and the `exact' shell corrections is found to
be indeed small, the maximum deviation being of the 
order 150 keV, implying that the factor $\alpha$ obtained merely by using the 
spherical nuclei works very well for deformed systems as well (with both 
deformation parameters $\beta_2$ and $\beta_4$). This observation is indeed of 
great practical importance.

\begin{table}[htb]
\caption{Values of the liquid drop parameters obtained through the
$\chi^2$ minimisation for `exact' and `approximate' shell corrections 
denoted by `E' and `A' respectively.}
\begin{center}
\begin{tabular}{|c|c|c|c|} \hline
Quantity      &  `E'     &    `A' \\ \hline
 $a_v$        & -15.435  & -15.421 \\ \hline 
 $k_v$        &  -1.875  &  -1.873 \\ \hline 
 $a_s$        &  16.673  &  16.580 \\ \hline 
 $k_S$        &  -2.430  &  -2.432 \\ \hline 
 $a_{cur}$    &   3.161  &   3.295 \\ \hline 
 $r_0$        &   1.219  &   1.221 \\ \hline 
 $C_4$        &   0.963  &   0.953 \\ \hline 
 $w_1$        &  -2.763  &  -2.652 \\ \hline
 $w_2$        &   3.725  &   3.659 \\ \hline
{\it rms}     &   0.610  &   0.607 \\ \hline 
\end{tabular}
\end{center}
\end{table}

With these approximate shell corrections, we make a re-fit to the liquid drop
parameters. Comparison between the liquid drop parameters as reported in 
Section 3 and the ones obtained with the approximate shell corrections 
is presented in Table 3.
It is indeed gratifying to note that the liquid drop parameters obtained 
in the two cases are almost identical, and so is the {\it rms} deviation 
of the calculated binding energies with respect to experiment \cite{WAP.03}. 
This substantiates the validity of the parametrisation of $\alpha$.

To test the robustness of the parametrisation of $\alpha$ further, we calculate 
the constants $\alpha _j$'s in Eq. (\ref{alpPH}) using just four nuclei 
($^{40}$Ca, $^{100}$Sn, $^{146}$Gd and $^{208}$Pb) instead of 2171 nuclei as 
described above.  It is found that the numerical values of the constants 
practically remain the same. To test the validity of these parameters, the 
liquid drop parameters are re-worked employing the new values of $\alpha_j$'s. 
It is found that the liquid drop parameters thus obtained are practically 
equal to the ones reported in the right most column of Table 3.

We close this section, by concluding that the absorption of fourth - order 
Wigner - Kirkwood corrections into the second contributions is reliable, and 
can be used in large scale mic-mac calculations. The absorption also has
the advantage of reducing the numerical noise that might arise in the higher
order derivatives of the potentials.

\section{Summary and Conclusions}
The semiclassical Wigner - Kirkwood $\hbar$ expansion of the one - body partition function
has been employed instead of the Strutinsky averaging scheme to calculate the 
shell corrections within the framework of a mic-mac
model. The microscopic part of the energy also contains pairing contributions 
that are obtained using the Lipkin - Nogami scheme. We have improved the 
macroscopic part of the model as compared with the one used in our previous 
work \cite{BHA.10,BHA.10a} by including the curvature term as well as the Wigner contribution. With 
just ten adjustable parameters, our model reproduces the binding energies of 
561 even - even spherical and deformed nuclei with {\it rms} deviation of 610 
keV. We have tested this new mic-mac model near the proton and neutron drip 
lines as well as in the superheavy region. Our present calculations show that
the mic-mac model proposed in this paper reproduces remarkably well the recent
experimental results in these exotic scenarios.

Further, a systematic study of the ratio of the fourth - order and second - 
order Wigner - Kirkwood energies has been carried out. We find that the ratio 
of these two energies behaves in a very systematic manner. We have shown that 
this ratio can be parametrised accurately by a simple expression, implying
that the fourth - order corrections can be absorbed into the second - order 
contributions in a very simple way. We have checked that using this simple
procedure, we recover practically the same parameters of the macroscopic part,
without deteriorating the quality of agreement achieved with the full Wigner
Kirkwood calculation including 
explicitly the fourth - order contributions. Therefore, this simplified 
calculation of shell corrections can be used confidently in the large scale 
mic-mac calculations that we plan to carry out as the next step.

Finally, we point out that there is still some room for improving our model
particularly 
in two specific directions. On the one hand, the full blocking procedure in
the pairing calculations of odd - odd, odd - even and even - odd nuclei,
that may be particularly relevant for spherical nuclei, has to be introduced.
On the other hand, refinements in the mean field Woods - Saxon potential 
and in the distance function are still needed to study with our model not only
neutron rich nuclei, but also fission barriers. This would require large scale
calculations with the model, for which, the simplification proposed above may
be very useful.

\begin{acknowledgments}
A.B. acknowledges partial financial support from Department of 
Science and Technology, Govt. of India (grant number SR/S2/HEP-34/2009).
M.C. and X.V. were partially supported by the Consolider Ingenio 2010
Programme CPAN CSD2007-00042, Grant No. FIS2011-24154 from MICINN and
FEDER, and Grant No. 2009SGR-1289 from Generalitat de Catalunya.

\end{acknowledgments}

\end{document}